 \documentclass[smallabstract,smallcaptions]{dccpaper}

\usepackage{cite}
\usepackage{graphicx}
\usepackage{url}
\usepackage{amsmath}
\usepackage{amssymb}
\usepackage[caption = false, font=footnotesize]{subfig}
\usepackage{float}
\usepackage{booktabs} 
\usepackage{multirow}
\usepackage[colorlinks=true,linkcolor=blue]{hyperref}%
\usepackage{array}
\usepackage{color}
\usepackage{tabularx}
\usepackage{longtable}
\usepackage{tabu}
\usepackage{pifont}
\usepackage{amssymb}
\usepackage{ragged2e}
\usepackage{tabularx}
\usepackage{longtable}
\usepackage{booktabs}
\usepackage{rotating}
\usepackage{makecell}
\begin{document}
\setcounter{secnumdepth}{4} 
\setcounter{tocdepth}{4} 

\newcommand\blfootnote[1]{%
  \begingroup
  \renewcommand\thefootnote{}\footnote{#1}%
  \addtocounter{footnote}{-1}%
  \endgroup
}

\title
{
\textbf{Beyond GFVC: A Progressive Face Video Compression Framework with Adaptive Visual Tokens}
}

\author{%
Bolin Chen$^{\ast}$,  Shanzhi Yin$^{\ast}$, Zihan Zhang$^{\ast}$, Jie Chen$^{\dag}$, Ru-Ling Liao$^{\dag}$, \\ Lingyu Zhu$^{\ast}$, Shiqi Wang$^{\ast}$  and Yan Ye$^{\dag}$\\[0.5em]
{\small\begin{minipage}{\linewidth}\begin{center}
\begin{tabular}{ccc}
$^{\ast}$City University of Hong Kong & \hspace*{0.5in} & $^{\dag}$DAMO Academy, Alibaba Group \\
\end{tabular}
\end{center}\end{minipage}}
}

\maketitle
\begin{abstract}
Recently, deep generative models have greatly advanced the progress of face video coding towards promising rate-distortion performance and diverse application functionalities. Beyond traditional hybrid video coding paradigms, Generative Face Video Compression (GFVC) relying on the strong capabilities of deep generative models and the philosophy of early Model-Based Coding (MBC) can facilitate the compact representation and realistic reconstruction of visual face signal, thus achieving ultra-low bitrate face video communication. However, these GFVC algorithms are sometimes faced with unstable reconstruction quality and limited bitrate ranges.
To address these problems, this paper proposes a novel Progressive Face Video Compression framework, namely PFVC, that utilizes adaptive visual tokens to realize exceptional trade-offs between reconstruction robustness and bandwidth intelligence. In particular, the encoder of the proposed PFVC projects the high-dimensional face signal into adaptive visual tokens in a progressive manner, whilst the decoder can further reconstruct these adaptive visual tokens for motion estimation and signal synthesis with different granularity levels. Experimental results demonstrate that the proposed PFVC framework can achieve better coding flexibility and superior rate-distortion performance in comparison with the latest Versatile Video Coding (VVC) codec and the state-of-the-art GFVC algorithms. The project page can be found at \url{https://github.com/Berlin0610/PFVC}.
\end{abstract}

\vspace{-0.6em}
\section{Introduction}
\vspace{-0.6em}
In recent years, visual data compression has been greatly powered by Artificial Intelligence Generated Content (AIGC) techniques, where the compression paradigms have been shifted from signal level~\cite{sullivan2012overview,bross2021overview} to generative representation~\cite{chen2024generative,icip2022zhao,chen2023csvt} or even semantic information~\cite{10032603,chen2023interactive}. Different from traditional hybrid-based coding frameworks, these generative video coding solutions have shown advantageous Rate-Distortion (RD) performance for visual content with strong priors and can support more diverse scenarios like metaverse-related or user-specified animation applications. In particular, these existing generative codecs mainly follow the early philosophy of Model-Based Coding (MBC)~\cite{7268565,lopez1995head} techniques and explore compact spatial/temporal representations to bridge the video compression \& generation tasks. 

Taking face videos as an example, they exhibit strong prior knowledge and statistical regularities, which can be easily incorporated into the Generative Face Video Compression (GFVC) frameworks~\cite{chen2024generative} towards high-efficiency visual communication. Different from early MBC techniques, these existing GFVC methods are mainly rooted in the latest face animation/reenactment models~\cite{FOMM,hong2022depth} and rely on the strong inference capabilities of deep models to achieve high-quality video reconstruction. More specifically, the GFVC encoder can characterize high-dimension visual signal into compact representations (like 2D keypoints~\cite{FOMM,konuko2021ultra}, 3D keypoints~\cite{wang2021Nvidia}, compact feature~\cite{chen2023csvt} and facial semantics~\cite{chen2023interactive}) for economical transmission costs via the analysis model. Furthermore, the GFVC decoder is capable of transforming these decoded compact information into fine-grained motion fields and employing the deep generative model to reconstruct realistic face signal. Although compact representations can facilitate promising RD performance within the GFVC pipelines, they also sometimes suffer unstable reconstruction quality and cover limited bitrate ranges, greatly limiting broader applications of GFVC. In other words, the compact representation technique is like a double-edged sword, which can greatly optimize signal representation costs but sometimes cause irreversible losses in signal reconstruction. In addition, this technique can only produce fixed-dimension representation and lack abilities to adapt to network bandwidth, leading to poor coding flexibility.

In view of these existing limitations, this paper proposes a novel Progressive  Face Video Compression framework based upon adaptive visual tokens, namely PFVC, aiming at facilitating generative face video communication towards bandwidth intelligence and robust reconstruction. In particular, inspired by the recent advancements of visual tokenization for the efficient signal synthesis~\cite{yu2024image}, the proposed PFVC framework transforms complex visual face signal into different-granularities visual motion tokens in a progressive manner. As such, these visual tokens are capable of hierarchically estimating facial motions and reconstructing the corresponding face contents according to different bandwidth scenarios. Moreover, the proposed progressive token strategy can allow these GFVC pipelines to be no longer limited to relying on compact representations for signal reconstruction, thus enhancing the effectiveness and robustness of the face generation process. The contributions of this paper are summarized as follows,

\begin{itemize}
\item We propose a novel PFVC framework that could progressively characterize high-dimensional visual face signal into adaptive sparse tokens and actualize different-quality face signal reconstruction. As such, the proposed PFVC scheme enjoys the advantages of coding flexibility, bandwidth intelligence and reliable reconstruction for face video communication.
\item We design an adaptive visual tokenization mechanism to describe facial motions at different granularities, such that the coded bitstream can be featured with different visual tokens in a progressive manner. Moreover, these adaptive visual tokens can be further transformed into pixel-wise dense motion fields for the temporal evolution inference via the implicit motion learning and GAN-based frame generation schemes.
\item We develop a progressive training strategy for generative video coding model, where different-granularities visual tokens are progressively introduced in each training stage and also guided with probability learning. Therefore, the proposed PFVC only needs one model to be compatible with overall bandwidth coverage instead of using multiple models.
\end{itemize}

\begin{figure*}[tb]
\vspace{-4.5em}
\centering
\centerline{\includegraphics[width=1\textwidth,height=6.5cm]{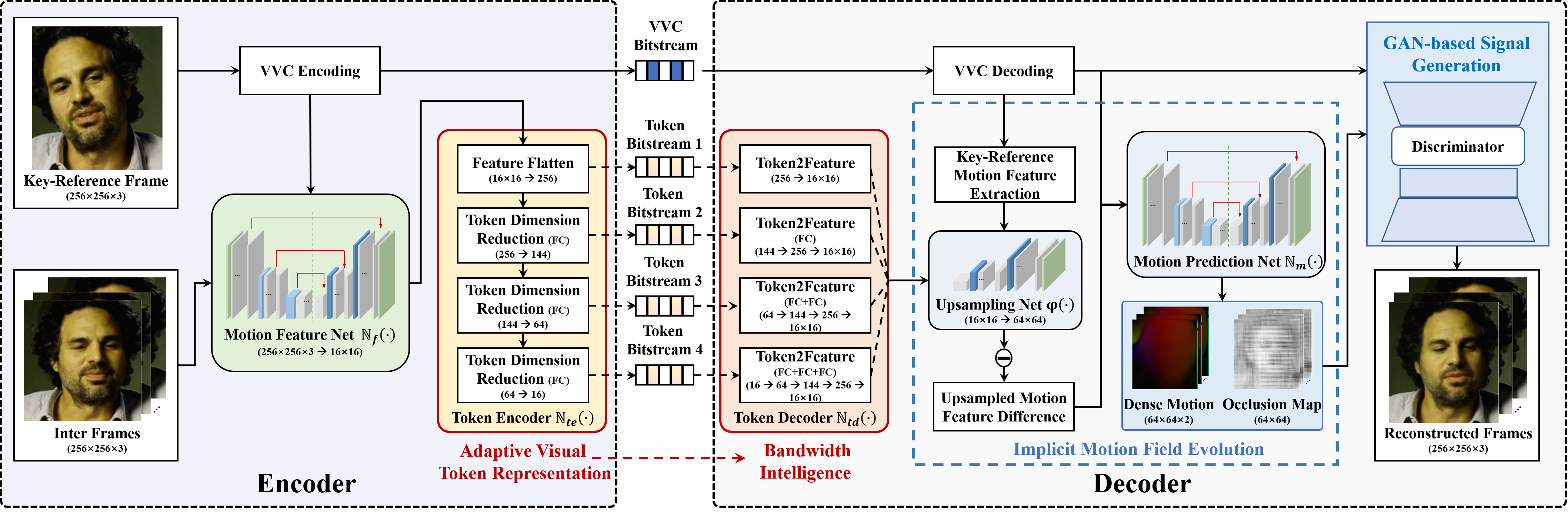}}
\caption{Overview of the proposed PFVC framework towards bandwidth intelligence.}
\label{fig1}
\vspace{-2em}
\end{figure*}

\vspace{-1.5em}
\section{The Proposed PFVC Framework}
\vspace{-0.6em}
In this section, we give an overall introduction regarding the proposed PFVC encoding/decoding processes. In addition, we discuss the design principle of each important component/module in the PFVC framework. Finally, we describe a new model training strategy for progressive video coding.  

\vspace{-0.6em}
\subsection{Overall Encoding/Decoding Processes}
\vspace{-0.2em}
As shown in Figure~\ref{fig1}, the proposed PFVC framework is mainly grounded on the adaptive visual tokens, thus exploring exceptional capabilities in the trade-offs between reconstruction robustness and bandwidth intelligence. At the encoder side, the input video signal can be classified as the key-reference frame $K$ (e.g., the first frame of this video) and subsequent inter frames $I_{l} \left (1\le l \le n , l\in Z  \right ) $.
The key-reference frame $K$ is intra-coded by an off-the-shelf VVC encoder~\cite{bross2021overview}, thus providing the rich texture reference for the subsequent frame reconstruction. Afterwards, the VVC-reconstructed key-reference frame $\hat{K}$ and these remaining inter frames $I_{l}$ are all characterized into a series of adaptive visual tokens in a progressive manner. Then, the network bandwidth scenario will further determine which granularity of visual tokens can be compressed. Then, these determined tokens are coded using inter prediction, quantization and entropy coding, where a context-adaptive arithmetic coder~\cite{TEUHOLA1978308,1096090} is employed to achieve optimal compression performance. 

The proposed PFVC decoder aims at using the decoded adaptive visual tokens to reconstruct face video at different quality levels. In particular, the off-the-shelf VVC decoder is employed to reconstruct the key-reference frame $\hat{K}$. In addition, the adaptive tokens of inter frames can be obtained by context-based entropy decoding and difference compensation. These decoded visual tokens are developed to infer the temporal motion evolution and transformed into pixel-wise dense motion fields, where their motion quality can be gradually improved with better token granularity. Finally, the powerful generative model can facilitate to reconstruct the face video with the guidance of the decoded key-reference frames and these inferred dense motion fields.

Thanks to the adaptive visual tokens, the proposed PFVC framework enjoys several  benefits that are not available to other existing GFVC algorithms. First, our proposed PFVC can characterize high-dimensional visual face signal into compact tokens with different granularities, ensuring greater feature adaptability in a learnable manner. In addition, different from the existing GFVC algorithms, which can only control the RD trade-offs by adjusting Quantization Parameter (QP) levels of the key-reference frame, our proposed PFVC can also achieve coding flexibility and bandwidth intelligence by transmitting visual tokens of different granularities. Furthermore, finer-granularity tokens can improve the face reconstruction robustness and stability, facilitating to solve the inaccurate motion estimation and unreliable generation problems that sometimes plague the existing GFVC pipelines due to compact representations. Finally, our proposed PFVC only needs one model to output visual features of different granularities and achieve face reconstruction of different qualities instead of using multiple GFVC models, which greatly expands the application possibilities.

\vspace{-0.6em}
\subsection{Key Techniques Introduction}
\vspace{-0.2em}
Herein, we will introduce all key techniques in the proposed PFVC framework, including adaptive visual token representation, implicit motion field evolution and GAN-based signal reconstruction. These designed schemes have greatly empowered the inference capability of the PFVC towards promising RD performance and bandwidth intelligence.

\textbf{Adaptive Sparse Tokens Representation.} 
These high-dimensional visual signal $\hat{K}$ or $I_{l}$ can be transformed to adaptive sparse tokens with different granularities via a specifically-designed token extractor. In particular, these face frames first execute down-sampling operation $\phi\left ( \cdot \right )$ through a scale factor $s$, and then fed into the feature network $\mathbb{N}_{f}\left ( \cdot \right )$ that is mainly composed of a U-Net architecture and a series of convolution/normalization operations to obtain motion features $\mathcal{F}_{motion}$ with the size of $1\times16\times16$. The motion features entailing accurate head posture and facial expressions can be learned in an unsupervised manner, 
\begin{equation}
\vspace{-0.7em}
{
 \mathcal{F}_{motion}= \mathbb{N}_{f}\left ( \phi \left ( X,s \right ) \right ) 
},
\end{equation} where $X$ could represent the VVC-reconstructed key-reference frame $\hat{K}$ or the inter frames $I_{l}$.

Afterwards, these learned motion features are further fed into a token network $\mathbb{N}_{t}\left ( \cdot \right )$ that is essentially an encoder-decoder translator with three Fully-Connected (FC) layers. Specifically, at the encoder part $\mathbb{N}_{te}\left ( \cdot \right )$  of this token network, the 2-D motion features are first flatten into a 1-D tokens with the dimension of 256, and then are sequentially sampled into three other 1-D tokens via FC layers with the dimensions of 144, 64 and 16. It should be noted that the number of FC layers and the specific dimension of each FC layer can be randomly set and herein we just provide an implementation example. The produced adaptive token list $\mathcal{L}_{tokens} $ is denoted as,   
\begin{equation}
\vspace{-0.7em}
{
\mathcal{L}_{tokens} = \mathbb{N}_{te}\left ( \mathcal{F}_{motion} \right ) 
},
\end{equation} where the dimension of adaptive tokens $\mathcal{L}_{tokens}$ is 256, 144, 64 and 16. In other words, depending on the given bandwidth environment, different-granularities visual tokens can be further coded into bitstream and actualize variable bitrate. After decoding these tokens $\mathcal{\hat{L}}_{tokens}$, they are input into the decoder part $\mathbb{N}_{td}\left ( \cdot \right )$ of the token network for the corresponding reconstruction of motion features $\mathcal{\hat{F}}_{motion}$,
\begin{equation}
\vspace{-0.7em}
{
\mathcal{\hat{F}}_{motion}= \mathbb{N}_{td}\left (\mathcal{\hat{L}}_{tokens} \right ) 
}.
\end{equation}
Overall, the proposed token network can realize feature sparsity and produce visual tokens of different granularities, thus providing great possibilities to explore compression scalability. 

\textbf{Implicit Motion Field Evolution.} Given the decoded key-reference frame $\hat{K}$, the extracted motion features $\mathcal{F}_{motion}^{\hat{K}}$ from $\hat{K}$ and the decoded motion features $\mathcal{\hat{F}}_{motion}^{I_{l}}$ from $I_{l}$, we design an implicit motion field evolution scheme to achieve  effective motion transformations from feature to pixel-wise fields. 
Different from CFTE~\cite{CHEN2022DCC}/CTTR~\cite{chen2023csvt} that both use the sparse-to-dense motion estimation strategy to generate a coarse deformed frame and then transform this frame for motion fields, our proposed scheme can directly use finer-granularity feature to infer motion fields via neural network learning. 
In particular, an up-sampled learning network $\varphi \left ( \cdot  \right )$ is employed to perform scale transformation for $\mathcal{F}_{motion}^{\hat{K}}$ and $\mathcal{\hat{F}}_{motion}^{I_{l}}$. In addition, difference between these scaled features is calculated,
\begin{equation}
\vspace{-0.6em}
{
Diff_{\left \langle I_{l},\hat{K} \right \rangle}=\varphi \left (\mathcal{\hat{F}}_{motion}^{I_{l}}\right )-\varphi \left (\mathcal{F}_{motion}^{\hat{K}} \right )
}.
\end{equation}
As an implicit motion change information, $Diff_{\left \langle I,K \right \rangle}$ is treated as the input together with the down-sampled $\hat{K}$ into the motion prediction network $\mathbb{N}_{m}\left ( \cdot \right )$, aiming at estimating pixel-wise dense motion map $\mathcal M^{I_{l}}_{dense}$ and occlusion map $\mathcal M^{I_{l}}_{occlusion}$ for face signal reconstruction. The process can be formulated as follows,
\begin{equation}
\mathcal M^{I_{l}}_{dense}=P_{1}\left (\mathbb{N}_{m}\left ( concat\left (\hat{K}, Diff_{\left \langle I_{l},\hat{K} \right \rangle} \right ) \right )  \right ),
\end{equation}
\begin{equation}
\mathcal M^{I_{l}}_{occlusion}=P_{2}\left (\mathbb{N}_{m}\left ( concat\left (\hat{K}, Diff_{\left \langle I_{l},\hat{K} \right \rangle} \right ) \right )  \right ),
\end{equation} where $P_{1}\left (  \cdot \right )$ and $P_{2}\left (  \cdot \right )$ represent two different predicted operation.

\textbf{GAN-based Signal Reconstruction.}
Analogous to the existing GFVC generator designs, we also adopt the flow-warped GAN module $\mathbb{N}_{GAN}\left ( \cdot \right )$ to reconstruct realistic face signal $\hat{I}_{l}$. In particular,
$\hat{K}$ is first warped with $M^{l}_{dense}$ for motion guidance and then the transformed features are further marked with $\mathcal M^{I_{l}}_{occlusion}$ for occlusion impainting,
\begin{equation}
{
\hat{I_{l}}= \mathbb{N}_{GAN}\left ( \mathcal M^{I_{l}}_{occlusion} \odot  f_{w}\left ( \hat{K} , \mathcal M^{I_{l}}_{dense} \right ) \right )
},
\end{equation} where $f_{w}$ and $\odot$ describe the back-warping and the Hadamard product operations. 

\vspace{-0.6em}
\subsection{Progressive Model Training Strategy}
\vspace{-0.2em}

\begin{table*}[]
\vspace{-3em}
\renewcommand\arraystretch{1.02}
\caption{Hyper parameter settings of the proposed progressive model training strategy} 
\label{table1}
\vspace{-0.5em}
\resizebox{\textwidth}{!}{
\begin{tabular}{ccccc}
\hline
Basic Settings                                                                                  & Stages S & Epoch N      & Granularity G          & Probability P                \\ \hline
\multirow{5}{*}{\begin{tabular}[c]{@{}c@{}}N=200\\ M=4\\ G={[}16, 64, 144, 256{]}\end{tabular}} & $S_{1}$       & 1$\sim$40    & {[}256{]}              & {[}1.0{]}                    \\
                                                                                                & $S_{2}$        & 41$\sim$80   & {[}144, 256{]}         & {[}0.7, 0.3{]}               \\
                                                                                                & $S_{3}$        & 81$\sim$120  & {[}64, 144, 256{]}     & {[}0.5, 0.3, 0.2{]}          \\
                                                                                                & $S_{4}$        & 121$\sim$160 & {[}16, 64, 144, 256{]} & {[}0.45, 0.25, 0.15, 0.15{]} \\
                                                                                                & $S_{5}$        & 161$\sim$200 & {[}16, 64, 144, 256{]} & {[}0.25, 0.25, 0.25, 0.25{]} \\ \hline
\end{tabular}
}
\vspace{-0.5em}
\end{table*}

To better facilitate the proposed PFVC to be compatible with different token granularities and achieve high-quality motion transformations  within the same model, we propose a new progressive model training strategy with the granular probability guidance. In particular, assuming that the PFVC model needs to be trained for $N$ epochs and can process $M$ granularities $G=\left \{ G_{1},G_{2},\dots,G_{M}  \right \} $, the overall model training will be divided into $S=\left \{S_{1},S_{2},\dots,S_{\frac{M}{N+1}}  \right \}, \frac{M}{N+1}\in Z$ stages. In each stage, the PFVC model will be assigned how to learn different token granularities with the given probabilistic ratios $P=\left \{ P_{1},P_{2},\dots,P_{M}  \right \} $. It should be mentioned that in each stage, the given probabilistic ratios for all granularities should be summed to 1. More importantly, to prevent the overall model capacity from being directly reduced by the coarse-grained tokens, the proposed progressive strategy prioritizes training of fine granularity in the early training stages and progressively introduce coarse-grained tokens in the subsequent stages. Besides, during the $S_{1} \sim  S_{\frac{M-N-1}{N+1}}$ stages, the maximum probability will be assigned to the corresponding newly-introduced token granularity to improve the PFVC model's acceptance of this granularity. As for the $S_{\frac{M}{N+1}}$ stage, the probability of each granularity will be set at the same ratio for stabilizing model training. 

In addition to the proposed progressive training strategy, the perceptual loss $\mathcal L _{per}$, adversarial loss $\mathcal L _{adv}$ and feature matching loss $\mathcal L _{fea}$ are adopted to supervise all PFVC stages in an end-to-end manner.
 The overall training loss can be described as,
\begin{equation}
\label{eq_training}
\begin{array}{c}
\mathcal L_{total}=\lambda _{per}\mathcal L _{per} + \lambda _{adv}\mathcal L _{adv} + \lambda _{fea}\mathcal L _{fea},
\end{array}
\end{equation} where $\mathcal L_{per}$ and $\mathcal L_{adv}$ are used to ensure the fidelity and naturalness of the reconstructed frames $\hat{I}_{l}$, and $\mathcal L _{fea}$ is introduced to stabilize the the GAN model training. 
In addition, the hyper-parameters are set as follows: $\lambda _{per}=10$, $\lambda _{adv}=1$ and $\lambda _{fea}=10$.

\vspace{-0.6em}
\section{Experimental Results}
\vspace{-0.6em}
This section first introduces the experimental settings, and then provides performance comparisons in terms of objective RD results and subjective quality.

\begin{figure*}[t]
\centering
\vspace{-5.5em}
\subfloat[Rate-DISTS]{\includegraphics[width=0.33 \textwidth]{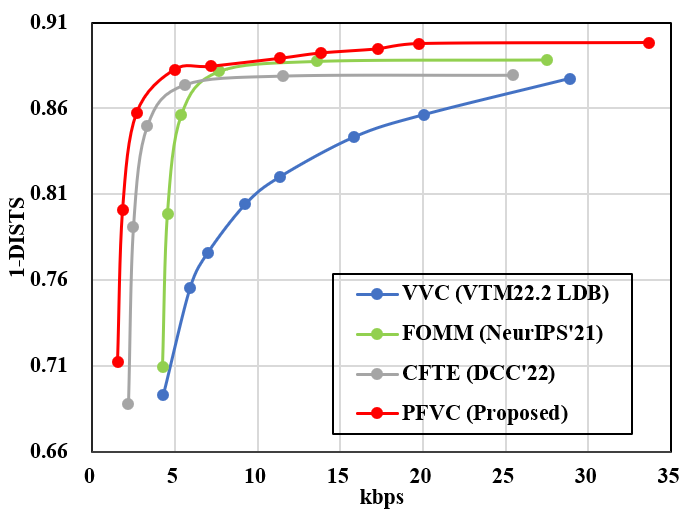}}
\subfloat[Rate-LPIPS]{\includegraphics[width=0.33 \textwidth]{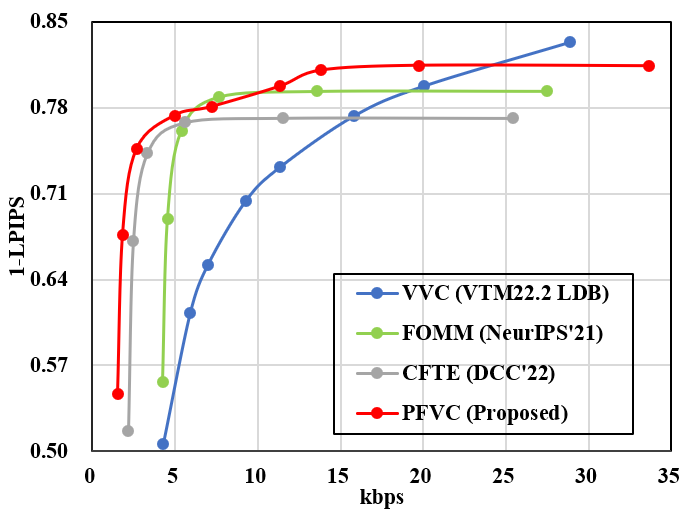}}
\subfloat[Rate-FVD]{\includegraphics[width=0.33 \textwidth]{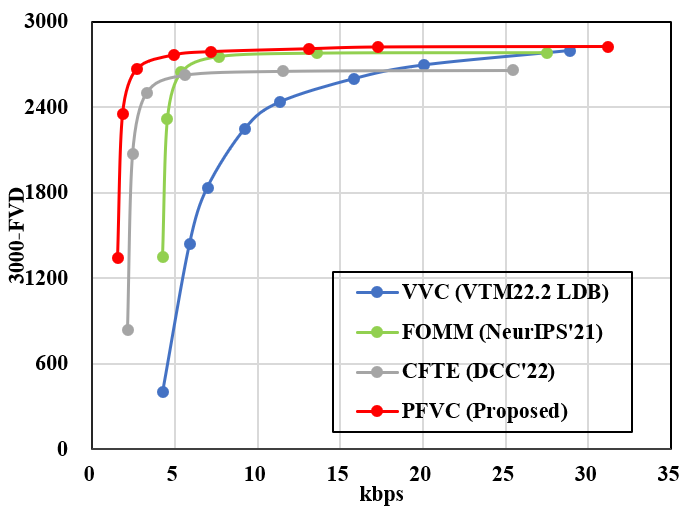}}\\
\vspace{-1em}
\subfloat[Rate-MANIQA]{\includegraphics[width=0.33 \textwidth]{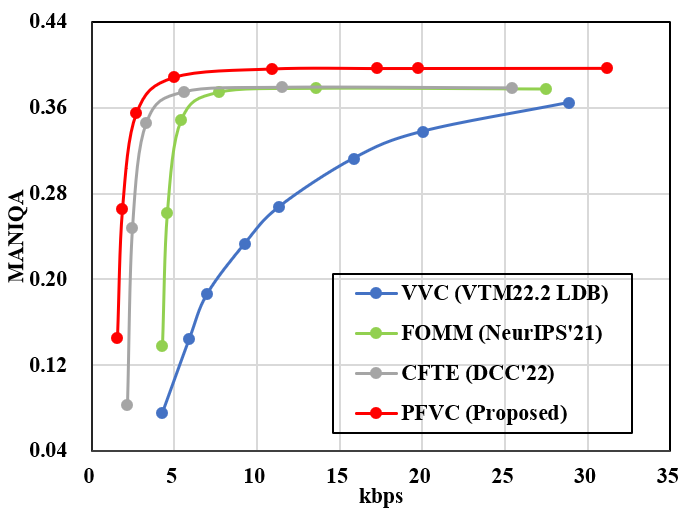}}
\subfloat[Rate-PSNR]{\includegraphics[width=0.33 \textwidth]{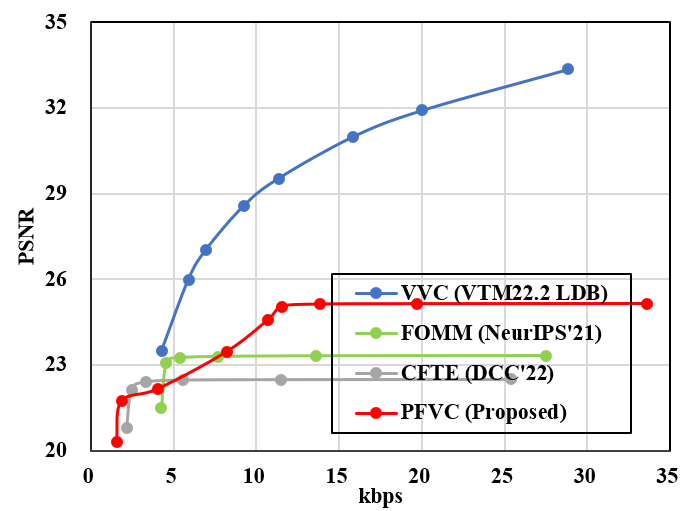}}
\subfloat[Rate-SSIM]{\includegraphics[width=0.33 \textwidth]{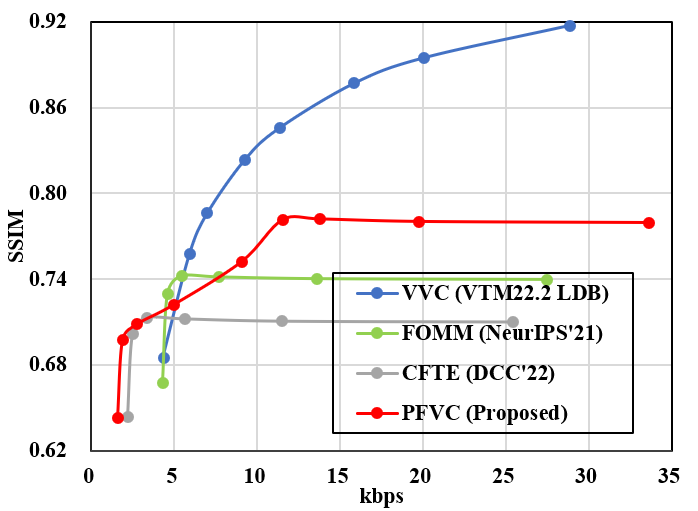}}
\caption{Overall RD performance comparisons with VVC~\cite{bross2021overview}, FOMM~\cite{FOMM} and CFTE~\cite{CHEN2022DCC} in terms of DISTS, LPIPS, FVD, MANIQA, PSNR and SSIM. }
\label{fig_RD}  
\vspace{-0.8em}
\end{figure*}

\vspace{-0.6em}
\subsection{Experimental Settings}
\vspace{-0.2em}
~~\quad\textbf{Implementation Details.}
We implement our proposed PFVC framework with Pytorch libraries, and train this model with the popular VoxCeleb training dataset~\cite{Nagrani17} on 2 NVIDIA TESLA A100 GPUs. During the model training phase, the Adam optimizer is used with the parameters $\beta _{1}$ = 0.5 \& $\beta _{2}$= 0.999, and the learning rate is set at 0.0002 for model convergence. In addition to the synchronized BatchNorm initialization and the data repeating strategy, we also employ the progressive training strategy as shown in Table \ref{table1}. 

\textbf{Compared Algorithms.}
To demonstrate the effectiveness of the proposed PFVC, we compare the state-of-the-art video coding standard VVC~\cite{bross2021overview} and two representative GFVC algorithm (CFTE~\cite{CHEN2022DCC} and FOMM~\cite{FOMM}). In particular, the testing dataset  adopted in~\cite{chen2023interactive} includes 50 talking face videos from the VoxCeleb testing dataset~\cite{Nagrani17}, where each sequence contains 250 frames with the resolution of 256$\times$256. 
The specific compared algorithm settings are provided as follows,
\begin{itemize}
\item{\textbf{VVC anchor:} is configured with the Low-Delay-Bidirectional (LDB) mode in VTM reference software 22.2, where the Quantization Parameters (QPs) are set at 32/35/37/40/42/45/47/52 and the coded signal format is YUV 4:2:0.}
\item{\textbf{CFTE/FOMM:} follows experimental testings specified in  the JVET GFVC AhG test conditions~\cite{JVET-AG2035}. In particular, the key-reference frame is converted to the YUV 4:2:0 format and compressed via the VTM reference software 22.2 with the QPs of 2/12/22/32/42/52. The subsequent inter frames can be represented into compact parameters and then compressed via a context-adaptive arithmetic coder.}
\item{\textbf{PFVC (Proposed):} Its test settings are basically the same as CFTE/FOMM except the feature representation of inter frames. Specifically, the proposed PFVC can characterize the inter frames into 4 different token granularities with the dimension of 16/64/144/256. Herein, the convex hull is performed on $6\times4=24$ RD points for a wider bitrate range and the optimal performance.}
\end{itemize}

\textbf{Evaluation Measures.} Herein, we adopt 4 learning-based quality measures and 2 traditional quality measures to evaluate the quality of reconstructed face video, including DISTS~\cite{dists}, LPIPS~\cite{lpips}, FVD~\cite{Unterthiner2019FVDAN}, MANIQA~\cite{yang2022maniqa}, PSNR~\cite{2009Mean} and SSIM~\cite{wang2004image}. These 6 adopted measures mainly involve perceptual-level, temporal consistency, no-reference GAN-based and pixel-level evaluations, where smaller scores of DISTS/LPIPS/FVD and higher scores of MANIQA/PSNR/SSIM indicate better perceived quality.
Therefore, Figure \ref{fig_RD} uses 1-DISTS, 1- LPIPS and 1-FVD  in order to present more coherent RD curves.
In addition to reconstruction quality, the coding bitrate (i.e., kbps) is also provided to further evaluate the compression performance.

\subsection{Performance Comparisons}

~~\quad\textbf{Rate-Distortion Performance.} Figure \ref{fig_RD} shows the RD performance of the VVC codec, FOMM, CFTE and our proposed PFVC scheme in terms of DISTS, LPIPS, FVD, MANIQA, PSNR and SSIM. It can be seen that the proposed PFVC can outperform the VVC codec for DISTS/LPIPS/FVD/MANIQA quality evaluations at overall bitrate ranges and for PSNR/SSIM quality evaluations at relatively low bit rate ranges. In addition, compared with the CFTE/FOMM anchors, the proposed PFVC can achieve the highest performance for all metrics and effectively improve the reconstruction quality with similar bit consumption.

As for pixel-level measurement, the proposed PFVC and other GFVC algorithms cannot achieve any advantages compared with the VVC codec. This is largely due to the fact that generative compression algorithms aim at reconstructing face signal in perceptual-level feature domain instead of pixel-level domain, which cannot adapt to the pixel-level evaluations (PSNR/SSIM)~\cite{10477607}. More importantly, by introducing finer token granularity, the proposed PFVC can gradually improve the GFVC ability in pixel-level reconstruction. Although the performance can be  improved, the adopted token granularities are still insufficient for the motion information supplementation, resulting in the saturation of its pixel-level reconstruction quality after a certain rate point.

\begin{figure}
\centering
\vspace{-5.2em}
\centerline{\includegraphics[width=1\textwidth]{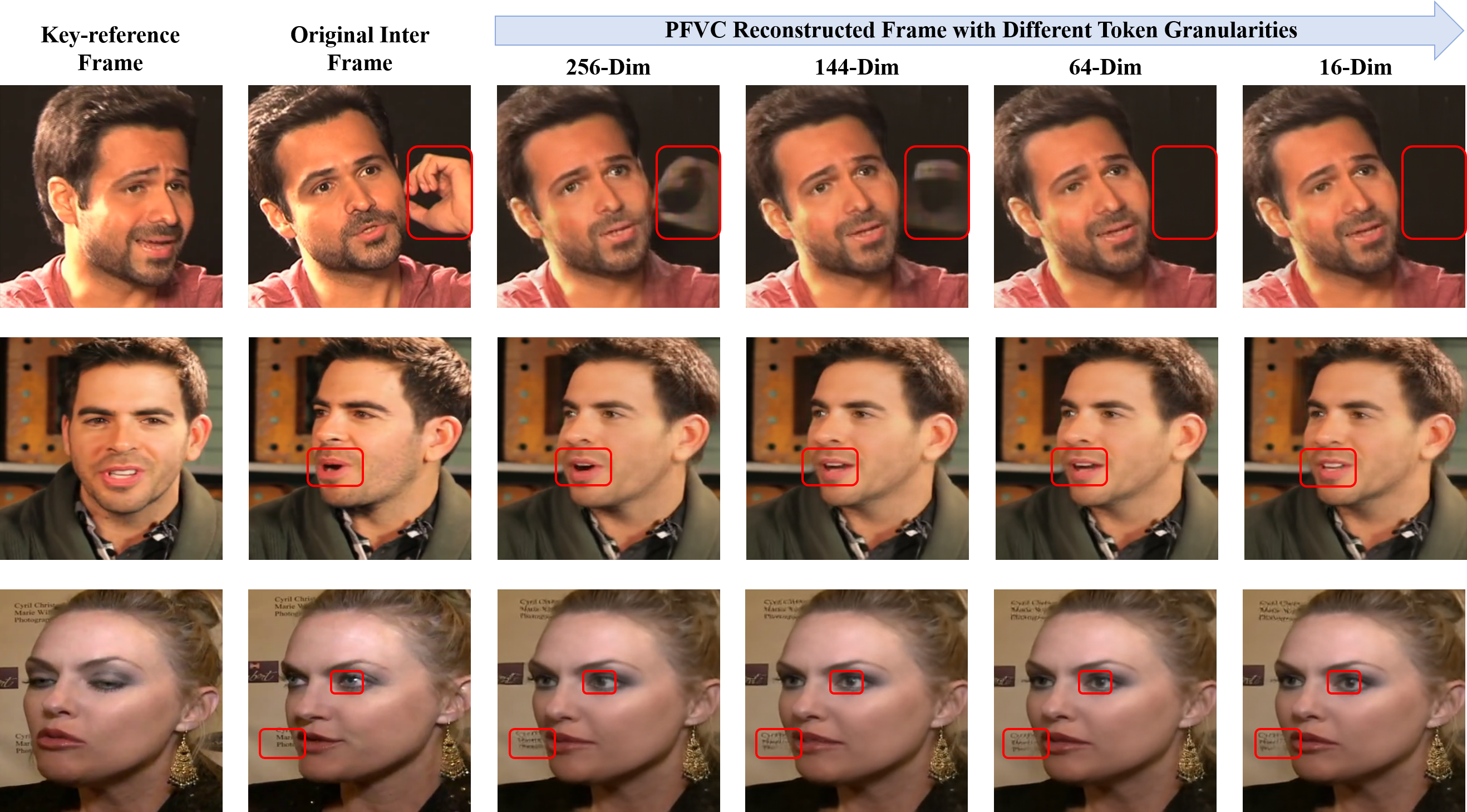}}
\caption{Subjective comparisons of reconstruction quality using the proposed PFVC with 4 different token granularities. More visual examples can be found in \href{https://github.com/Berlin0610/PFVC}{project page}. } 
\label{fig:demo_progressive} 
\vspace{-1em}
\end{figure}

\begin{figure*}[t]
\centering
\vspace{-4.5em}
\subfloat[Visual Example 1]{\includegraphics[width=0.68\textwidth]{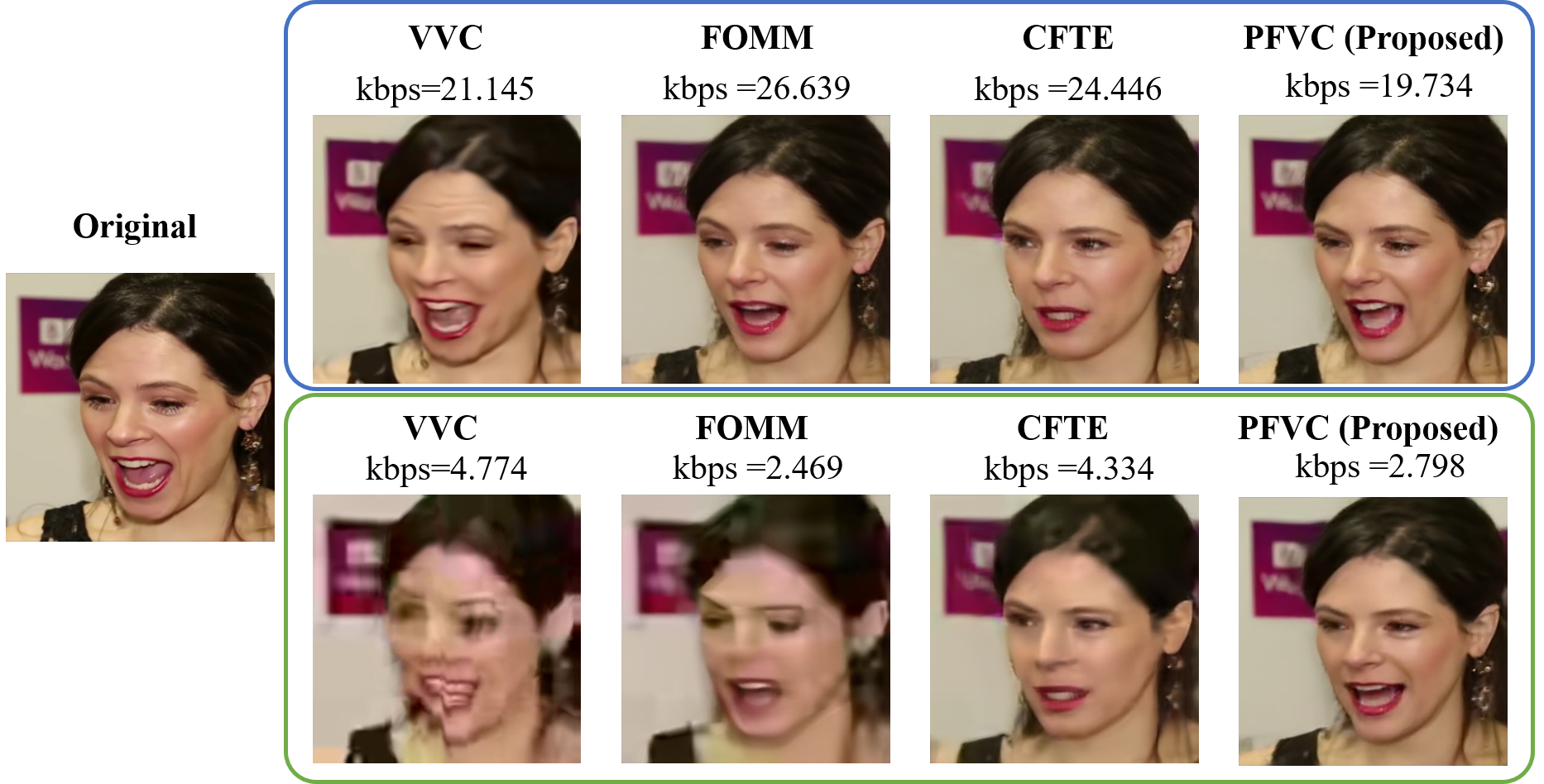}}\\
\vspace{-1em}
\subfloat[Visual Example 2]{\includegraphics[width=0.68 \textwidth]{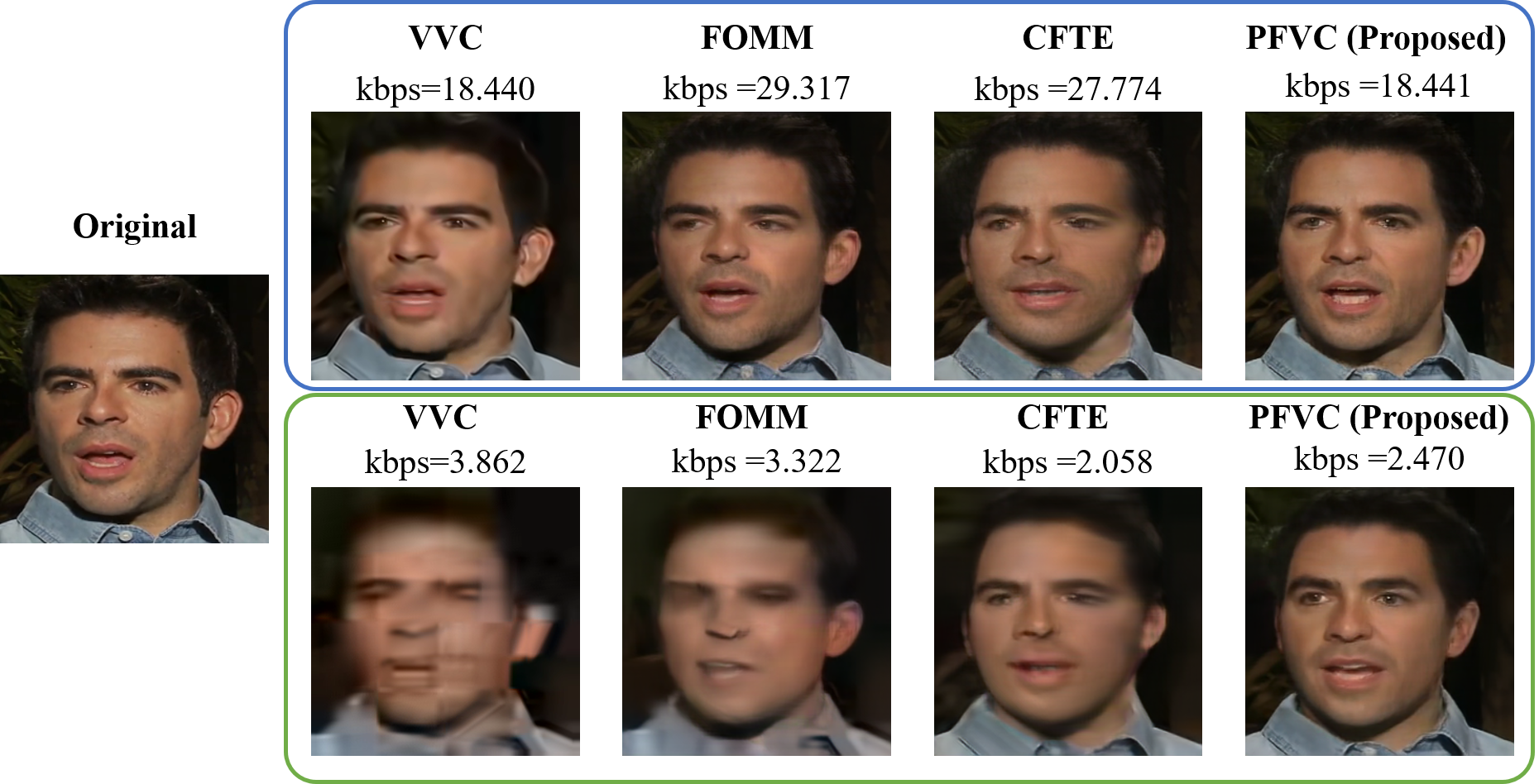}}\
\caption{Subjective comparisons of reconstruction quality using different compression algorithms like VVC~\cite{bross2021overview}, FOMM~\cite{FOMM} and CFTE~\cite{CHEN2022DCC} and our proposed PFVC at 2 different bitrate levels. More visual examples can be found in \href{https://github.com/Berlin0610/PFVC}{project page}. }
\label{fig_RD}  
\vspace{-0.8em}
\end{figure*}

\textbf{Subjective Quality.}
Figure~\ref{fig:demo_progressive} provides visual examples to verify that the proposed PFVC framework can reduce some annoying visual artifacts or recover more missing details via different-granularities visual token representations. For examples, with the provision of finer token granularity, it is possible to infer the hand details that originally do not exist in the key-reference frame. In addition, the finer token granularity can stabilize the background generation and facilitate higher face fidelity such as accurate mouth movement and eye direction.  

As illustrated in Figure~\ref{fig_RD}, the proposed PFVC algorithm is able to reconstruct face frames with high signal fidelity and accurate facial motion. In particular, at very low bit rates, the VVC reconstructed results exhibit  blocking artifacts, while the proposed PFVC still can reconstruct vivid face signal. Besides, the facial motions in reconstructed video quality using FOMM and CFTE are not accurate, while the proposed PFVC can infer more precise head posture and facial expression.

\vspace{-0.6em}
\section{Conclusion}
\vspace{-0.6em}
This paper proposes a novel progressive generative face video compression framework to actualize coding flexibility, bandwidth intelligence and reliable reconstruction for face video communication. In particular, adaptive visual tokens are leveraged to describe facial motion changes in a progressive manner, whilst the deep generative model is employed to learn implicit motion fields from these visual tokens and perform temporal inference for high quality face reconstruction. Experimental results demonstrate that compared with the latest VVC codec and other GFVC algorithms, our proposed PFVC can achieve more promising RD performance and wider overall bitrate as well as quality converge.

\vspace{-0.7em}
\section*{References}
\vspace{-0.6em}
\bibliographystyle{IEEEtran}
\bibliography{main}
\vspace{-0.8em}
\end{document}